\documentclass{elsart}
\def\ohalf{{\textstyle{1\over 2}}}

\def\half{{\textstyle{1\over 2}}}
\def\vqhalf{{\textstyle{\vec{Q}\over 2}}}

\def\osix{{\textstyle{1\over 6}}}

\def\tthird{{\textstyle{2\over 3}}}
\def\fivehalf{{\textstyle{5\over 2}}}

\newcommand{\beq}{\begin{equation}}
\newcommand{\be}{\begin{equation}}
\newcommand{\eeq}{\end{equation}}
\newcommand{\ee}{\end{equation}}
\newcommand{\beqa}{\begin{eqnarray}}
\newcommand{\bea}{\begin{eqnarray}}
\newcommand{\eeqa}{\end{eqnarray}}
\newcommand{\eea}{\end{eqnarray}}
\newcommand{\bra}[1]{\langle {#1} |}                        
\newcommand{\ket}[1]{| {#1} \rangle}

\usepackage{graphicx}
\usepackage{bm,hyperref,color}
\usepackage{epsfig}

\begin{document}
\begin{frontmatter}
\title{Electromagnetic form factors of pion and rho in
the three forms of relativistic kinematics}

\author{Jun He$^{1,3}$,}
\author{B. Juli\'a-D\'{\i}az$^2$, and}
\author{Yu-bing Dong $^1$}%

\address{ $^1$Institute of High Energy Physics, Chinese
Academy of Sciences,P.O.Box 918-4,100039 Beijing, P.R.China\\
$^2$Helsinki Institute of Physics and Department of Physical
Sciences, P.O.Box 64, 00014 University of Helsinki, Finland\\
$^3$Graduate School of the Chinese Academy of Sciences,
\\ 100039, Beijing, P.R.China}

\date{4 July 2004}

\begin{abstract}
The electromagnetic form factors of the $\pi$ and the $\rho$ are
obtained using the three forms of relativistic kinematics, instant
form, point form and (light) front form. Simple representations of
the mass operator together with single quark currents are employed
with all the forms. The Poincar\'e covariant current operators are
generated by the dynamics from single-quark currents that are
covariant under the kinematic subgroup. Front and instant forms allow 
to reproduce the available data for the pion form factor. 
On the other hand point form is not able to reproduce qualitatively 
the experimental data with reasonable values for the wave function 
parameters. For the $\rho$ electromagnetic form factors, instant and 
front forms provide a consistent picture. The obtained results do not 
depend appreciably on the wave function used.
\end{abstract}
\end{frontmatter}

The electromagnetic form factors of hadrons are an important
source of information about their internal structure. They 
provide a useful tool to understand the dynamics of the strong 
interaction and the role played by relativity in understanding 
the transition region between the low-energy and perturbative QCD 
domains.

In the literature, there are several works where the form factors 
of the $\pi$ and the $\rho$ have been studied making use of relativistic 
quark models, e.g. 
Refs.~\cite{Chung:mu,keister,cardarelli,Amghar:2003tx,Allen,Krutov}. 
Most theoretical studies were carried out making use of front form 
while only lately point form was also employed 
giving rise to some discrepancies in its formulation~\cite{Amghar:2003tx,Allen}.
Here we present a comparative study of the form factors obtained with 
the three forms of relativistic kinematics making use of the same 
assumptions for the mass operator and the structure of the electromagnetic 
current. The understanding of the different formulations of relativistic 
quark models and their ability to provide a coherent picture of hadrons 
with simple assumptions is of interest as it can serve as a framework to 
understand all the new data on the $Q^2<$ 12 GeV$^2$ region. 
Our study aims at exploring the advantages and drawbacks of the different formulations.

The study of form factors making use of relativistic quark models requires 
a relation between the variables which enter in the representation of the 
mass operator, $\vec k_i$ and spins, and the variables which enter in the 
vertex and appear in the current. The relation between these two sets of 
variables depends on the ``form of kinematics'' being used. The three forms 
are named as point, instant and front form. They differ from each other in 
the kinematical subgroup of the Poincar\'e group. In point form the kinematical 
subgroup is the full Poincar\'e group, in instant kinematics it is the group of 
rotations and translations at a fixed time, while in front form it is the 
group that leaves invariant the light cone.

Electromagnetic form factors of two-body systems can be defined
as certain matrix elements of the electromagnetic current. In point and 
instant forms, the charge form factor of $S=0$ mesons can be defined as 
follows,
\begin{eqnarray}
F_C(Q^2)=\langle 0, {\vec Q}/2|I^0(0)|0,-{\vec Q}/2\rangle_c
\end{eqnarray}
where $I^0$ is the time component of the current and ${\vec Q}$ has been taken 
to be parallel to the $z$-axis.

In front form, in the $Q^+=0$ frame, the charge form factor can be extracted 
from the ``plus'' component of the current, $I^+=n\cdot I$, with $n=\{-1,0,0,1\}$:
\begin{eqnarray}
F_C(Q^2)=\langle0|I^+(0)|0\rangle \, ,
\end{eqnarray}
in this case the momentum transfer is taken to be transverse to the
$z$-direction~\cite{Riska04}.

For $S=1$ mesons, such as the $\rho$, we adopt the definition of 
Ref.~\cite{Chung}. For point and instant forms, we have:
\begin{eqnarray}
G_C(Q^2)&=&
{1\over3} \left [ \langle0,\vqhalf|I^0(0)|-\vqhalf,0\rangle_c 
+2\langle1,\vqhalf|I^0(0)|-\vqhalf,1\rangle_c \right], \nonumber\\
G_M(Q^2)&=&\sqrt{2\over\eta}
\langle1,\vqhalf|I_+(0)|-\vqhalf,0\rangle_c,\nonumber\\
G_D(Q^2) &=&\frac{1}{2\eta}
\left[\langle0,\vqhalf|I^0(0)|-\vqhalf,0\rangle_c
-\langle 1,\vqhalf|I^0(0)|-\vqhalf,1\rangle_c \right],
\label{eq:rhoFFpi}
\end{eqnarray}
while for front form,
\begin{eqnarray}
&&G_{C}(Q^2)=F_{0d}+\osix F_{2d}-\tthird\eta \left\{F_{0d}+F_{2d}+\fivehalf F_{1d}\right\},\nonumber\\
&&G_{M}(Q^2)=2F_{0d}+F_{2d}+F_{1d}(1-\eta),\nonumber\\
&&G_{D}(Q^2)=\frac{1}{\eta}\left\{F_{2d}+\eta\left(\ohalf F_{2d}-{F_{0d}-F_{1d}}\right)\right\}.
\label{eq:rhoFFf}
\end{eqnarray}
where
\begin{eqnarray}
&&F_{0d}(Q^2)=\frac{1}{2(1+\eta)}\{\langle1|I^+(0)|1\rangle+\langle0|I^+(0)|0\rangle\},\nonumber\\
&&F_{1d}(Q^2)=\frac{-\sqrt{2}}{\sqrt{\eta}(1+\eta)}\langle1|I^+(0)|0\rangle,\nonumber\\
&&F_{2d}(Q^2)=\frac{-1}{(1+\eta)}\langle1|I^+(0)|-1\rangle \, .
\label{eq:Fi}
\end{eqnarray}
The kinematical variable $\eta$ is defined as $\eta={1\over 4}(v_f-v_a)^2=Q^2/4M^2$, 
where $M$ is the meson mass. In a previous work~\cite{Allen}, the momentum 
appears scaled as $p=\frac{M}{2m_q}\frac{Q}{2}$, which means,
$\eta=\frac{Q^2}{16m_q^2}$, where $m_q$ is the mass of quark. 

With the definitions in Eqs.~(\ref{eq:rhoFFpi}),~(\ref{eq:rhoFFf})
and~(\ref{eq:Fi}), the charge and magnetic and quadrupole moments
of spin 1 mesons are defined as,
\begin{eqnarray}
eG_C(0)=e,\ \ eG_M(0)=2M\mu,\ \ eG_Q(0)=M^2{\mathcal D},
\label{quadrupole}
\end{eqnarray}
where $e$ is the electron charge and $M$ is the meson mass.

Meson states are represented by eigenfunctions of the mass operator, which 
are functions of internal momenta, $\vec k_i$, and spin variables. We use a 
simple spectral representation of the mass operator, considering only the $\pi$ 
and the $\rho$. The meson wave functions are constructed in the naive quark 
model~\cite{Close},
\begin{eqnarray}
\psi^\pi({\vec q}) &=& \xi_c \, \varphi_{0}({\vec q}) \, \phi_S
\,\chi_A,
\nonumber \\
\psi^\rho({\vec q})&=& \xi_c \, \varphi_{0}({\vec q}) \,\phi_A \,
\chi_S,
\end{eqnarray}
where $\xi_c$ is the fully symmetric color wave function. The flavor wave 
functions $\phi_{S,A}$ have the forms: 
\beqa
\phi_{S,A}^{+} &=&\frac{1}{\sqrt{2}}(u\overline{d}\pm\overline{d}u)\, , \nonumber \\
\phi_{S,A}^{0} &=&\frac{1}{2}[(d\overline{d}-u\overline{u})
\pm(\overline{d}d-\overline{u}u)] \, , \nonumber \\
\phi_{S,A}^{-} &=&-\frac{1}{\sqrt{2}}(d\overline{u}\pm\overline{u}d) \,.
\eeqa
The spin wave functions, $\chi$, are the usual:
\beqa
\chi_S^1=\uparrow \uparrow \, , \quad 
\chi_{S,A}^0 &=& {1\over \sqrt{2}}(\uparrow \downarrow \pm \downarrow\uparrow) \, ,\quad
\chi_S^{-1}=\downarrow \downarrow \, .
\eeqa

The effect of the Lorentz transformation on the spin variables for canonical 
spins is accounted by a Wigner rotation of the form:
$
D^{1/2}_{\lambda_i,\sigma_i}\left(R_W[B(v_K),k_i]\right)$
with
$
R_W[B(v_K),k_i]:= B^{-1}(p_i)B(v_K)B(k_i)\, ,
$
where $B(v)$ are rotationless Lorentz transformations, and $v_K$ is the 
boost velocity.

For the spatial part of the wave function, we adopt both Gaussian
and rational forms: 
\beq 
\varphi^G_{0}({\vec q})=\frac{1}{(b\sqrt{\pi})^{3/2}}e^{-{\vec q}^2/2b^2}, \qquad
\varphi^{R}_{0}({\vec q})={\mathcal N}(1+{\vec q}^2/2b^2)^{-a} \,, 
\eeq
where ${\vec q}=\frac{1}{\sqrt{2}}({\vec k}_2-{\vec k}_1)$ and
${\mathcal N}$ is a normalization constant. In the center of mass
frame we have ${\vec k}_1+{\vec k}_2=0$ and thus 
$\vec k_2 = {1\over \sqrt{2}} \vec q = -\vec k_1$. 
The Jacobians of the transformation between the variables are: \\ 
for point form, 
\beq 
J(\vec v;\vec p_2):=
\left({\partial \vec q \over \partial \vec p_2}\right)_{\vec v} =
2\sqrt{2}{ \omega_2\over E_2}=  2\sqrt{2}{(E_2 v^0-p_{2z}v_z)\over
E_2}\, . 
\label{pointjac} 
\eeq 
for front form 
\beq 
J({\bf P};{\bf
p_2}):= \left({\partial \vec q \over \partial  (\xi_2, {\bf
k}_{2\perp})} \right)_{\bf P} =  2\sqrt{2}{\partial k_z \over
\partial  \xi}=2\sqrt{2}\frac{M_0}{4\xi(1-\xi)}\, , 
\eeq 
with
\beqa 
k_{zi}&=&{1\over2}\left(\xi_i M_0-{m_q^2+k_{i\perp}^2\over
\xi_i M_0}\right)
=M_0(\xi-1/2) \,,  \nonumber \\
M_0^2&=& \sum_i{m_q^2+k_{i\perp}^2\over \xi_i}={m_q^2+k_{\perp}^2\over \xi(1-\xi)}.
\eeqa
and for instant form,
\beq
J(\vec P,\vec p_2) =2\sqrt{2}\frac{\omega_2}{E_2}\left\{1-\frac{E_2v_z}{M_0}\left(
{p_{1z}\over E_1}-{p_{2z}\over E_2}\right )\right\}, \label{instjac}
\eeq
where
\beq
P_x=P_y=0\; ,
\;
M_0^2= (\sum _i E_i)^2-|\vec P|^2 \; ,
\;
\vec v := {\vec P\over M_0}\,.
\eeq

For each form of kinematics the dynamics generates the current density
operator from a kinematic current. For point form we have,
\bea
&&\bra{\vec v_f,\vec
v_2'}I^\mu(0) \ket{\vec v_2,\vec v_a}= 
\delta^{(3)}(v_2'-v_2)(\osix+\half \tau_3^{(1)})\bar u(\vec
v_1\,') \gamma^{(1)\mu}u(\vec v_1)\, ,
\label{cur4}
\eea
for front form,
\beqa
&&\bra{P^+, P_{\perp f},{\bf p}_2'}I^+(x^-,x_\perp)
 \ket{{\bf p_2}, P_{\perp a},P^+} \nonumber \\
&&=\delta^{(3)}(p_2'-p_2)(\osix+\half \tau_3^{(1)}) \bar u({\bf
p_1'}) \gamma^{(1)+}u({\bf p_1}) e^{\imath( P_{\perp f}- P_{\perp
a})\cdot  x_\perp}\, ,
\eeqa
and for instant form,
\beqa
&&\bra{\half \vec Q,\vec p'_2}I^\mu(\vec x) \ket{\vec p_2,-\half \vec Q } \nonumber \\
&&=
\delta^{(3)}(p_2'-p_2)(\osix+\half \tau_3^{(1)})\bar u(\vec
p_1\,')\gamma^{(1)\mu}u(\vec p_1) e^{\imath(\vec Q\cdot \vec x)}\,.
\eeqa

With the formulas given above, we can calculate the form factors
of the $\pi$ and the $\rho$. The procedure used to fix the
meson states is the following. We fix $a$, $b$ and $m_q$ (or just
$b$ and $m_q$ for the gaussian case) so that they are both in the
range of other similar calculations and that the $\pi$ form factor
and charge radius are fairly reproduced. The use of two different
wave functions allows us to estimate the
theoretical uncertainty derived from the wave function used.

The first relevant issue we notice is that it is not possible
to find a set of parameters with any of the wave functions
in point form so that the $Q^2$ behavior of the form factor is
reproduced. This was one of the points raised in Ref.~\cite{Amghar:2003tx}.
For instant and front forms it is possible to find such a set of 
parameters using both types of wave functions. The sets of 
parameters are given in Table~\ref{parameters}.

\begin{table}[t]
\begin{center}
\begin{tabular}{ l cccc}
\hline
\hline
                   &b [MeV]    &$m_q$[MeV]& a &  $\sqrt{\langle r^2_\pi\rangle}\ [$fm$]$\\
\hline
\hline
\multicolumn{5}{c}{Gaussian}\\
\hline
Instant form         &370        &  140    &  $--$    &0.600\\
Point form           &3000       &  380    &  $--$    &3.018\\
Front form           &450        &  250    &  $--$    &0.665\\
\hline
\multicolumn{5}{c}{Rational}\\
\hline
Instant form       &700        &  150    &5         &0.619\\
Point form         &3000       &  300    &1         &2.545\\
Front form         &600        &  250    &3         &0.659\\
\hline
\end{tabular}
\caption{Parameters and charge radius of the $\pi$ in instant,
point and front form both for the rational and gaussian spatial 
wave functions. The experimental value for the charge radius is
$\sqrt{\langle r^2_\pi\rangle}$=0.663$\pm0.006$ fm~\cite{Amendolia}.
\label{parameters}}
\end{center}
\end{table}

\begin{figure}[t]
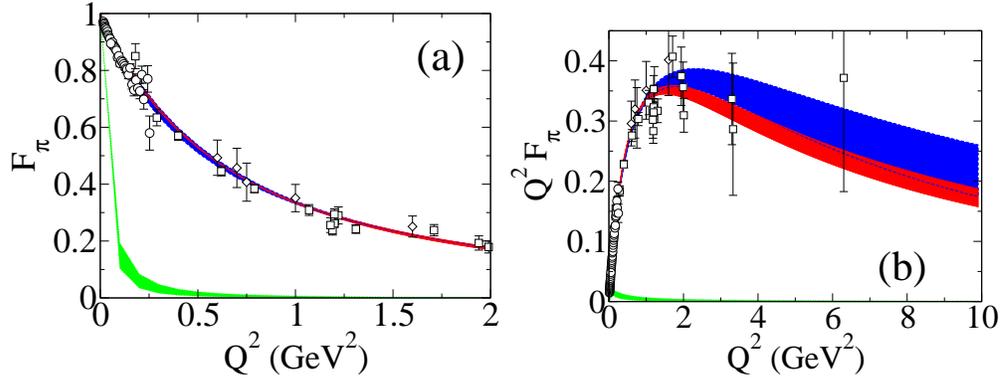

\vspace{20pt} \mbox{\epsfig{file=fig1a, width=65mm}}
\mbox{\epsfig{file=fig1b, width=65mm}} \caption{(a) $\pi$ charge
form factor as function of $Q^2$(GeV$^2$). The band is obtained as
explained in the text. Red, green and blue stand for
instant, point and front form. (b) Same as (a) but
multiplied by $Q^2$. The experimental data are from
Refs.~\protect\cite{Amendolia,Bebek,Volmer}} \label{pion}
\end{figure}

In Fig.~\ref{pion} the $\pi$ form factor obtained with the
parameters of Table~\ref{parameters} is presented. The bands 
depicted in the figures are constructed using the results 
obtained with the gaussian and rational wave functions, one  
gives the band minimum while the other provides the maximum. 
In this way the band gives an estimate of the theoretical uncertainty due
to the specific choice of wave function. The chosen parameter sets
permit a good reproduction of the $Q^2$ behavior of the data in instant 
and front forms. The charge radii calculated in front and 
instant forms are quite close to the experimental data, small 
discrepancies with the data could be attributed to our simple model 
were known effects arising from vector meson contributions to the charge 
radius of the pion are not accounted for~\cite{Bernard}.
The result obtained with point form is completely off and cannot be 
brought into agreement by changing the parameters of the model wave 
functions. 

The high $Q^2$ behavior of the form factor is qualitatively 
similar in instant and front form although the instant form result 
falls slower. In both cases the falloff of the form factor at large 
$Q^2$ is faster than the QCD predictions of Refs.~\cite{Farrar,Brodsky}, 
$Q^2 F(Q^2) \propto $ const or ($1/\log{Q^2}$). In fact the obtained behavior 
is closer to $Q^2 F(Q^2) \propto 1/Q^{2}$. This faster 
falloff, of almost one power of $Q^2$, seems to be a general 
trend in most quantum mechanical calculations where the coupling 
of the photon to the standard quark current is 
considered~\cite{privdesplanques}. Improvements, e.g. considering 
two-body currents or different quark-photon couplings, are beyond 
the scope of this letter.

We have shown that the $\pi$ form factor can be reasonably 
understood with instant and front form of kinematics by finding 
the appropriate mass operator, which in our case corresponded 
to finding the parameters of Table~\ref{parameters}. Now we consider 
the case of the $\rho$ meson. Due to the fact that the parameters for point
form could not be fixed from the pion charge form factor, we 
chose them as similar to those of Ref.~\cite{Allen} (let us note 
that for the $\rho$ the prescription used in their paper is irrelevant 
due to the fact that $\rho \approx 2 m_q$). 
In Table~\ref{rhotable} the values for the magnetic and quadrupole moments 
defined in Eq.~(\ref{quadrupole}) are 
presented. The results for the $\rho$ magnetic moment in all cases are 
smaller than 2 $e/2M_\rho$, and also smaller than other theoretical
estimates. The quadrupole moments obtained, which ranges between
$[0.2-0.5]\ e/M_\rho^2$, are consistent with Refs.~\cite{Choi,Jaus}.

\begin{table}[t]
\begin{center}
\begin{tabular}{ l cc}
\hline
\hline
                     & $\mu$ [$e/2M_\rho$] & ${\mathcal D}$  [$e/M_\rho^2$]  \\
\hline
Instant form         & 1.5                                    &  [0.36$-$0.29] \\
Point form           & 0.9                                    &  [0.38$-$0.50] \\
Front form           & 1.5                                    &  [0.2$-$0.33] \\
Choi {\it et al.} \protect\cite{Choi} & 1.9                    & 0.43 \\
Jaus \protect\cite{Jaus}              & 1.83                   & 0.33 \\
Cardarelli {\it et al.} \protect\cite{cardarelli}    & 2.23    & 0.61 \\
\hline
\end{tabular}
\caption{Magnetic and quadrupole moments of the $\rho$ for instant, point,
and front form. The range correspond to using gaussian or rational
wave functions.\label{rhotable}}
\end{center}
\end{table}

\begin{figure}[t]
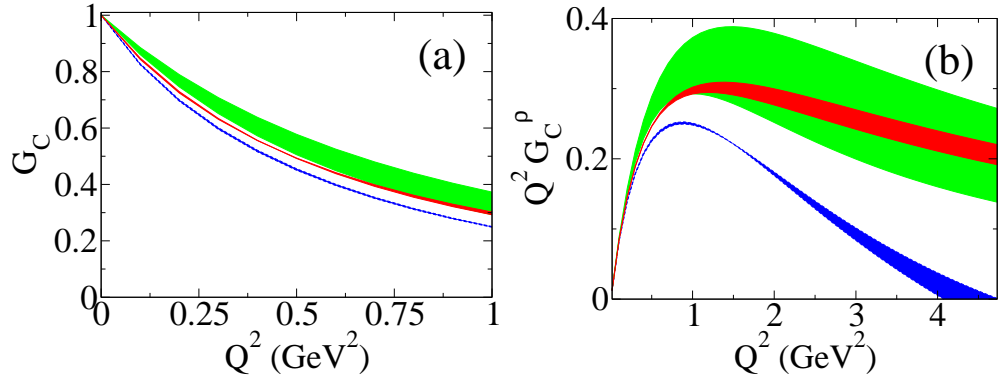

\vspace{20pt} \mbox{\epsfig{file=fig2a, width=65mm}}
\mbox{\epsfig{file=fig2b, width=65mm}} \caption{ Charge form
factor of the $\rho$ as function of $Q^2$(GeV$^2$). Same
description as Fig.~\ref{pion}.\label{crho}}
\end{figure}

\begin{figure}[t]
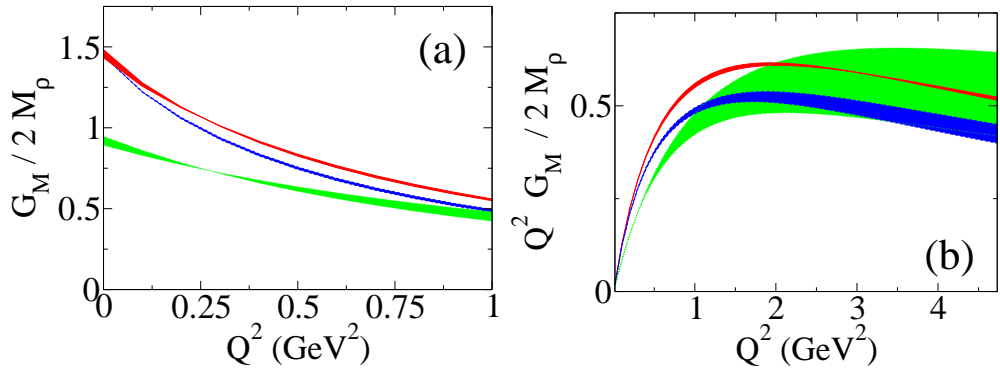

\vspace{20pt} \mbox{\epsfig{file=fig3a, width=65mm}}
\mbox{\epsfig{file=fig3b, width=65mm}} \caption{ Magnetic form
factor of the $\rho$ over $2M_\rho$ as function of $Q^2$(GeV$^2$).
Same description as Fig.~\ref{pion}.\label{mrho}}
\end{figure}

\begin{figure}[t]
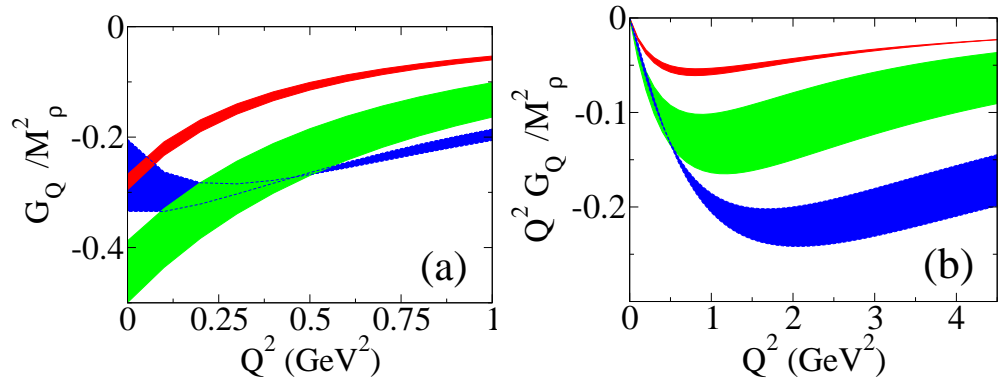

\vspace{20pt} \mbox{\epsfig{file=fig4a, width=65mm}}
\mbox{\epsfig{file=fig4b, width=65mm}} \caption{ Quadrupolar form
factor of the $\rho$ over $M_\rho^2$ as function of
$Q^2$(GeV$^2$). Same description as Fig.~\ref{pion}.\label{drho}}
\end{figure}

In Figs.~\ref{crho},~\ref{mrho} and~\ref{drho} we present the $Q^2$ dependence
of the electromagnetic form factors defined in Eqs.~(\ref{eq:rhoFFpi})
and~(\ref{eq:rhoFFf}).

In Fig~\ref{crho} the charge form factor is depicted. Unlike in
the case of the $\pi$ it can be seen that in this case the three
forms provide a coherent picture in the low $Q^2$ region.
However, in the high-$Q^2$ region the situation is different.
Essentially, point and instant form predict a behavior close to
the one observed in the pion charge form factor, while front 
form falls faster and eventually crosses zero at $Q^2\approx 4.5$ 
GeV$^2$. This feature of the charge form factor becoming negative 
in front form calculations is also present in the electric form factor 
of the proton, see Ref.~\cite{Riska04}, and in other front-form 
calculations of the $\rho$ charge form factors~\cite{keister,cardarelli}. 
The $Q^2$ dependence of the form factors at high $Q^2$ is mostly 
independent of the wave function used as can be easily seen by the 
thinness of the bands.

The failure of point form to reproduce the $\pi$ form factor
is therefore most likely due to the small mass of the $\pi$, as
explained in Ref.~\cite{Desplanques04}. On the other hand our study
shows that for higher mass mesons, such as the $\rho$, it is possible
to find an appropriate mass operator such that point form gives
qualitatively similar results to front or instant form, leaving the
case of the $\pi$ as a pathological one.

The magnetic form factor is shown in Fig.~\ref{mrho}. In this case
instant and front form predict a similar magnetic moment, which is
given in Table~\ref{rhotable}, while point form predicts a
magnetic moment which is 30\% lower. The high $Q^2$ behavior is
similar for the three forms and is compatible with $\propto
1/Q^2$.

Similar situation, but in this case with point form providing
a larger value, appears in the quadrupole form factor, which
is given in Fig.~\ref{drho}. Instant and front forms give
similar quadrupole magnetic moments although with a very different
prediction for the $Q^2$ dependence of the form factor.

We have studied the electromagnetic form factors of the
$\pi$ and the $\rho$ making use of the three different
forms of relativistic kinematics. Front and instant forms 
provide a correct picture of the $\pi$ electromagnetic form factor, 
giving both the correct charge radius and $Q^2$ dependence.
The high-$Q^2$ dependence predicted for the charge form 
factor of the pion is faster than the one predicted from 
QCD calculations. Both front and instant forms give similar results for 
the electromagnetic form factors of the $\rho$. 
Point form does not allow a description of the $\pi$ electric 
form factor, most likely due to its small mass, and, although 
qualitatively similar, gives different quantitative values 
for the $\rho$ electromagnetic form factors. Our calculated values 
for the $\rho$ dipole and quadrupole moment are around 20\% smaller 
than the ones available in the literature. 

The authors want to thank F. Coester for valuable 
comments on the manuscript. This work is supported by the 
National Natural Science Foundation of China (No. 10075056 and No. 90103020), by
CAS Knowledge Innovation Project No. KC2-SW-N02. B. J.-D. thanks
the European Euridice network for support (HPRN-CT-2002-00311) and
the Academy of Finland through grant 54038.

\end{document}